\documentstyle[10pt]{article}
\newenvironment{mytitle}{\begin{center} \LARGE}{\\ [.1in]\end{center}} 
\newenvironment{myauthor}{\begin{center} }{\\ [.1in]\end{center}} 
\newenvironment{myinstit}{\begin{center} \it}{\end{center}}

\begin{document}

\begin{mytitle}
Understanding quantum scattering properties in terms of purely classical dynamics. 2D open chaotic billiards.
\end{mytitle}
\vspace{.5cm}

\begin{myauthor}
J. A. M\'endez-Berm\'udez$^1$, G. A. Luna-Acosta$^1$, P. \v{S}eba$^{2,3}$, 
and K. N. Pichugin$^{2,4}$.
\end{myauthor}

\begin{myinstit}
$^1$Instituto de F\'{\i}sica, Universidad Aut\'onoma de Puebla, Apdo. Postal 
J-48, Puebla 72570, M\'exico.\\
$^2$Department of Physics, University Hradec Kralove, Hradec Kralove, Czech Republic.\\
$^3$Institute of Physics, Czech Academy of Sciences, Cukrovarnicka 10, 
Prague, Czech Republic.\\
$^4$Kirensky Institute of Physics, 660036 Krasnoyarsk, Russia.
\end{myinstit}

\begin{abstract}
We study classical and quantum scattering properties in the ballistic regime of particles in two-dimensional chaotic billiards that are models of electron- or micro- waveguides. To this end we construct the purely classical counterparts of the Scattering Probability (SP) matrix $\mid S_{n,m}\mid^2$ and Husimi distributions and specialize to the case of mixed chaotic motion. Comparison between classical and quantum quantities allows us to discover the purely classical dynamical origin of certain general, as well as particular features that appear in the quantum description of the system. On the other hand, at certain values of energy the tunneling of the wavefunction into classically forbidden regions produces striking differences between the classical and quantum quantities. We also see the manifestation of Whispering Gallery Orbits as a self-similar structure in the transmission part of the classical SP matrix. A potential application of this phenomenom in the field of microlasers is discussed briefly.\\


\end{abstract}

\vspace{.5cm}

\section{Introduction}

At present, the majority of studies on the Quantum-Classical Corresponcence (QCC) of chaotic systems concern {\it bounded} motion, for which very important results have been obtained \cite{Felix,Linda,Stockmann}. Most works also treat the situation when chaos is {\it fully} developed. Here we would like to contribute to the understanding of the QCC for {\it open} systems with {\it mixed} chaotic motion, in particular we consider particle motion in two-dimensional (2D) billiards whose phase space is characterized by incomplete (Smale) horse-shoes. As is well known, 2D billiards are popular models of ballistic mesoscopic systems \cite{mesos,ketz2,Akguc} and microwave cavities \cite{miccav}.\\

Usually and naturally, QCC is explored by means of semiclassical calculations \cite{JalaBara}. Unfortunatelly, this approach poses limitations since semiclassical quantities involve the cumbersome determination of the action associated with the trajectories and sums which often do not converge. However, as Baranger and co-workers \cite{JalaBara} have found through their semiclassical calculations of conductance, the dominant contributions to the average quantum conductance are classical. This motivates us to construct purely classical counterparts of the Scattering Probability (SP) matrix $\mid S_{n,m}\mid^2$ and Husimi distributions for 2D open billiards. In a previous paper \cite{gala5} we showed that knowledge of the classical counterpart of the SP matrix enables us to predict the global structure of the quantum SP matrix. Here we further our study, specializing to the case of mixed chaotic motion, by: 1) analyzing in more detail the type of trajectories responsible for the formation of certain structures in the classical and quantum SP matrices; 2) comparing Husimi distributions with classical phase space structures; and 3) identifying certain purely quantum effects in the behaviour of the conductance.\\

This paper is organized as follows. In Section II we review the definition
of the SP matrix and the construction of its classical counterpart. In Section 
III the 2D waveguide model is presented, as well as its resulting quantum and classical SP matrices for a large number of modes. Section IV concerns the dynamical analysis of the classical SP matrix, the existence of whispering gallery orbits, and the structure that these create in the classical SP matrix. Some aspects of the QCC are studied in Section V in terms of Husimi distributions and their classical counterparts, the {\it Transient Poincar\`{e} Maps}. The appearance of purely quantum effects in our system is analyzed in Section VI and in Sect. VII we make some concluding remarks.\\

\section{Quantum and Classical SP matrices}

{\bf Quantum SP matrix}\\

Quantum scattering is studied via the $S$ matrix, $\hat{\bf S}$, which relates incoming to outgoing waves,

\begin{equation}
V^{out} = \hat{\bf S}\ V^{in},
\end{equation}

\noindent where the $V^{in}$ and $V^{out}$ vectors specify, respectively, waves coming into and going out of the interaction region. In the case of a 2D waveguide of arbitrary shape connected to two leads, say, left (L) and right (R) leads, the wavefunction in the leads are of the form

\begin{eqnarray}
\Psi^L (x,y) = \sum_{m=1}^{M_L} \left[ a_m^L \exp(ik^L_mx) + b_m^L
\exp(-ik^L_mx)\right] \phi^L_m(y) \nonumber \\
\Psi^R (x,y) = \sum_{m=1}^{M_R} \left[ a_m^R \exp(ik^R_mx) + b_m^R
\exp(-ik^R_mx)\right] \phi^R_m(y)
\end{eqnarray}

\noindent where

\begin{equation}
\phi^{L,R}_m(y)= \sqrt{\frac{2}{d_{L,R}}} \sin \left( 
\frac{m\pi y}{d_{L,R}} \right) 
\end{equation}

\noindent is the part of the wavefunction associated with the motion perpendicular to the direction of propagation; $d_L$ ($d_R$) is the constant width of the left (right) lead, and $M_L$ ($M_R$) is the total number of propagating modes supported by the left (right) lead at a given Fermi energy $E$. For the rest of the paper we shall consider $d_L=d_R=d$, and hence 
$M_L=M_R=M$.\\

Note that with this notation the $S$ matrix and the incoming and outgoing wave vectors take the form 

\begin{eqnarray}
\hat{\bf S} = \left( \begin{array}{ll} t & r' \\ r & t' \end{array} 
\right ),\ \ V^{in} = \left( \begin{array}{l} a^L \\ b^R \\ \end{array}
\right),\ \ V^{out} = \left( \begin{array}{l} a^R \\ b^L \end{array}
\right),
\end{eqnarray}

\noindent where $t$ ,$t'$, $r$, and $r'$ are the transmission and reflection $M \times M$ matrices. $M$ is the highest mode (the largest $m$ beyond which the longitudinal wave vector 
$k^{L,R}_m =\sqrt{\frac{2E}{\hbar^2}-\frac{m^2 \pi^2}{d_{L,R}^2}}$ becomes 
complex). The symbols $a^{L,R}$ and $b^{L,R}$ stand for the vectors 
$a^{L,R}_m$ and $b^{L,R}_m$, $m=1$, $2 \ldots M$. The elements of the 
transmission ($t$ and $t'$) and reflection ($r$ and $r'$) matrices are the transmission and reflection amplitudes given, respectively, by $t_{mn}(E)=(\sqrt{k_n/k_m})a_n^R/a_m^L$,
$t_{mn}'(E)=(\sqrt{k_n/k_m})b_n^L/b_m^R$,
$r_{mn}(E)=(\sqrt{k_n/k_m})b_n^L/a_m^L$, and
$r_{mn}'(E)=(\sqrt{k_n/k_m})a_n^R/b_m^R$. 
The {\it squared} modulo element $\mid t_{n,m}\mid^2$ ($\mid t'_{n,m}\mid^2$) 
gives the probability amplitude for a left (right)-incoming mode $m$ to  be transmitted to the right (left) lead into the mode $n$. Similarly, $\mid r_{n,m}\mid^2$ ($\mid r'_{n,m}\mid^2$) is the probability for a left (right)-incoming mode $m$ to be reflected to the left (right) lead into mode $n$. \\

The Scattering Probability (SP) matrix is defined by the $2M$ by $2M$ matrix  $\mid S_{n,m}\mid^2$, representing the transition probability for the incoming mode $m$ to transmit or reflect into an outgoing mode $n$.\\

{\bf Classical SP matrix}\\

Given the expression for the energy in the leads

\begin{equation}
E=\frac{\hbar^2}{2m_e} \left( k_m^2 +\frac{m^2 \pi^2}{d^2} \right),
\end{equation}

\noindent where $k_m$ and $\frac{m\pi}{d}$ are, respectively, the longitudinal and transversal components of the total wave vector {\bf K} with magnitude $K = \sqrt{2m_eE}/\hbar$, the angle $\theta_m$ between $k_m$ and $K$ is given by

\begin{eqnarray}
\theta_m = \sin^{-1} \left[ \frac{m \pi}{d} \frac{1}{K} \right] = \sin^{-1} \left[ \frac{m \pi \hbar}{d\sqrt{2m_eE}} \right].
\end{eqnarray}

Classically a particle can enter the lead at any angle in the continuous range $-\frac{\pi}{2} < \theta < \frac{\pi}{2}$, but since the $M$ open modes for a given energy $E$ are discrete, we associate a range of angles $\Delta \theta_m\equiv\theta_m-\theta_{m-1}$ to each mode $m$. That is, we coarse-grain the classical angles. Clearly the classical limit is $M=\infty$.\\

We now explain the procedure to construct the purely classical counterpart of $\mid S_{n,m} \mid^2$. Consider a classical particle entering the cavity of the waveguide, say, from the left lead and making an angle $\theta_i$ within a range corresponding to a given mode $m$. The particle (ray) will generally collide with the walls of the cavity of the waveguide a few times before exiting the cavity either to the left or to the right lead, making a certain angle $\theta_f$ to which we can associate a mode $n$ if $\theta_f \in \Delta \theta_n$. To specify the initial conditions for the trajectory of the i-{\it th} particle, the initial position $(x_i,y_i)$ and the initial angle $\theta_i$ must be given. In order to account for all possible types of trajectories, we take a large number (typically $10^5$) of initial positions for each incoming angle $\theta_i$. By recording the number of particles scattered into the various ranges of $\theta$ associated with different outgoing modes ${n}$, we obtain a distribution of outgoing modes for each incoming mode $m$. This distribution gives the {\it classical counterpart} of the matrix elements $\mid t_{n,m} \mid^2$ and $\mid r_{n,m} \mid^2$ of the quantum SP matrix. Similarly, to obtain the classical counterpart of 
$\mid t'_{n,m} \mid^2$ and $\mid r'_{n,m} \mid^2$ we repeat the above process but for particles entering from the right lead. This defines the procedure for constructing the {\it classical SP matrix}.\\

\section{The System}

We chose the geometry of the cavity of the waveguide to be that of the ``rippled" billiard, which consists of two hard walls: a rippled wall 
modulated by a periodic function $y(x)=d+a\xi(x)$ [$\xi(x+L)=\xi(x)$], and a flat wall at $y=0$; $d$ is the width of the waveguide and $a$ the modulation parameter. For concretenes, the periodic function in this paper is given by $\xi(x)=1-\cos(2\pi x/L)$. To form the waveguide, we attach two semi-infinite collinear leads of width $d$ to a cavity whose length is an integer multiple of the period $L$. Fig. 1 shows the geometry for a two-period rippled cavity.\\
 
This {\it finite} length version of the rippled billiard, a model of a quantum or electromagnetic waveguide, has been used to study certain transport manifestations of chaos in the classical \cite{Licht,gala1} and quantum \cite{ketz2,Akguc} regimes. Moreover, studies of the {\it infinitely} long (periodic with no leads) rippled billiard, originally introduced to model beam acceleration problems \cite{Month}, has also provided insight in the understanding of general features of periodic structures (e. g., energy band structure, structure of eigenfunctions, etc.) and has been utilized to explore the problem of quantum-classical correspondence of classically chaotic systems \cite{gala2,gala3,gala4}.\\

Before analyzing our scattering system (cavity connected to leads) it is instructive to look at the dynamics of the infinitely long periodic ({\it i.e.}, $x$ mod. $L$) billiard. A Poincar\`{e} Map (PM) of the system supplies us with the dynamical panorama. We choose the Poincar\`{e} Surface of Section (PSS) as the bottom boundary, $y=0$ and the PM is generated by the iteration of the Birkhoff pair of variables $(x_j,\theta_j)$, labeling the longitudinal component of the position and the angle the total momentum of the particle makes with the $x$-axis right after its $j^{th}$ collision with the bottom wall. Previous studies \cite{ketz2,Licht,gala1,Month,gala2,gala3,gala4} have shown that the motion can be either regular, mixed, or fully chaotic depending on the relationship between the geometrical parameters $(d,a,L)$. A representative picture of a mixed phase space (obtained with the set of parameters $(d,a,L) = (1,0.305,5.55)$) is shown in Fig. 2.\\

In this work, we analyze the waveguide system defined by a cavity formed by a {\it single} period of the rippled billiard [with the same values $(d,a,L)$ used in Fig. 2] attached to two aligned semi-infinite leads each of width $d$. The period one and period four resonance island structure of Fig. 2 is formed by trapped orbits within the region $0<x<L$. In particular the central resonance is formed by trajectories colliding nearly perpendicular with the walls in the neighbourhood of $x=L/2$, the widest part of the cavity. Since the left (right) lead is attached to the cavity at $x=0$ ($x=L$), such island structure is also present in the waveguide system. Note that the orbits forming these islands are classically not accesible to scattering trajectories.\\

In Figures 3 and 4 we present, respectively, the quantum and classical SP matrix for the {\it one-period} waveguide with $M=200$ open channels. The quantum SP matrix was calculated by the recursive Green's function method \cite{ketz2}, and the classical SP matrix by the procedure described above with $10^3$ ensambles (each characterized by a different $\theta_i$) of $10^5$ different initial conditions. Note the rich structure present mainly in the transmission part ($\mid t_{n,m} \mid^2$ and $\mid t'_{n,m} \mid^2$) of the SP matrix. We note that $t = t'$ and $r = r'$, as expected from the symmetry of the system.\\

The similarity between quantum and classical SP matrices is remarkable in 
the case of a large number of modes (here $M=200$), suggesting that the motion belongs to the deep semiclassical regime. Nevertheles, we have also seen a resemblance between the quantum and classical SP matrices even for a number of modes as low as 10. Hence the calculation of the classical SP matrix as a tool for prediction becomes relevant in quantum regimes.\\

\section{Dynamical analysis of the classical SPM}

First we focus on the transmission part of the classical SP matrix which is presented in Fig. 5, this time as a function of angles. The first clear and interesting structure to analyze is in the region $0.345\leq \theta \leq 0.48$, for which an enlargement is made in Fig 6. The existence of {\it Whispering Gallery Orbits} (WGO) is well known in cavities with concave walls. These are guided orbits along the inner surface of such cavities, see e.g. \cite{NocSton}. Note that our cavity has a concave part in $L/4 < x < 3L/4$. In recent works, it has been shown that WGOs produce self-similar structures \cite{Kruelle,Seba}. Here we can see that a self-similar structure also shows up in the SPM, in particular in the transmission part of the classical SPM (Fig. 6).\\

We find that the structure of Fig. 6 is formed by trajectories that hit 
only on the concave part of the {\it upper} boundary before being transmitted: the outer part of the structure ($\theta > 0.465$) is due to trajectories that hit it only once; the second generation ($\theta \simeq [0.41,0.465]$) is due to trajectories that hit it twice, the next generation ($\theta \simeq [0.39,0.41]$) to trajectories that hit the boundary three times, and so on. The point of convergence of the self-similar structure (not shown in Fig. 5 due to numerical lack of resolution) is $\theta = 0.3325$, the angle at the inflection point of the profile ($x = L/4$). A particle that would hit the upper boundary at $x = L/4$ with $\theta_i = 0.3325$ would hit the concave part of the billiard an infinite number of times before leaving it at an angle $\theta_f = 0.3325$. In Fig. 7 whispering gallery trajectories selected from different generations of the structure of Fig. 6 are shown.\\

Similarly, we can discover the types of particle motion that produce the various  structures in the transmission part of the classical SP matrix. Fig. 8 shows typical trajectories that contribute to the zones marked in Fig. 5. All these orbits are very stable: small variations of the initial conditions $(x_i,y_i,\theta_i)$ follow closely the same trajectory, with the same number of collisions with the upper and lower boundaries and with very similar final angles $\theta_f$. Noting that the marix elements $\mid S_{n,m}\mid^2$ gives a measure of the number of trajectories connecting a pair of angles $\theta_i \rightarrow \theta_f$, then for {\it every high intensity structure in the classical SP matrix there corresponds a boundle of stable trajectories}. Moreover, since all the high intensity structures present in the classical SP matrix are also present in its quantum counterpart we can expect the stable orbits of the underlying classical motion to play the dominant role in determining quantum transport properties.\\

\section{Transient Poincar\`{e} Map (TPM)}

A complementary tool to analyze a waveguide system is the {\it Transient 
Poincar\`{e} Map} (TPM) which has the same meaning as the Poincar\`{e} Map (PM) but now taken for the scattering system in the range of interaction. The TPM (with surface of section at $y=0$) is presented in Fig. 9. Fig. 9a shows the part of the TPM generated by trajectories whose initial conditions start in the left lead only and Fig. 9b shows the contribution of trajectories starting from both, right and left leads. Comparison of Fig. 9b with Fig. 2 shows, as expected, that the stability regions present in the PM for the infinite billiard become forbidden phase space regions in the TPM.\\

Now we can ask about the location in the TPM of the whispering gallery orbits 
(WGO). For this purpose we construct the TPM choosing now as surface of section the {\it top} boundary because, as explained in Sect. IV, the WGOs are guided orbits along the concave part of the boundary. The upper TPM is constructed by the pairs $(x_j,\theta'_j)$ corresponding, respectively, to the longitudinal component of the position and the angle made by the total momentum with the tangent of the boundary at $x_j$, right after the $j^{th}$ collision with the upper wall. The location of the WGO in the TPM is shown (highlighted) in Fig. 10, where the numbers label the number of collisions the particles make with the upper boundary. A structure of this type in phace space formed by WGOs has been reported in \cite{Seba}. We see that this pattern evolves from top to bottom; the higher the number of collisions with the upper wall, the closer the angle $\theta'$ is to zero. In the limit $\theta'=0$, a particle will collide an infinite number of times along the concave part of the boundary.\\

A popular tool used to explore quantum-classical correspondence in billiards is the Husimi distribution \cite{hus}; it is the projection of a given quantum state onto a coherent state of minimum uncertainty. The Husimi distribution can be viewed as a quantum phase space probability density that can be directly compared with the classical phase space. See \cite{gala2} for details of the calculation of the Husimi distributions for the infinetly periodic rippled billiard. Here, for our waveguide system we calculate the Husimi distributions for each one of the $m$ opened modes at certain Fermi energy and compare them with the TPM generated solely by trajectories starting with the corresponding $\Delta \theta_m$. Then, using a Fermi energy that supports 20 modes we present in Figures 11a-11d the TPM's generated by particles entering from the left lead with angles corresponding to the $2^{nd}$, $5^{th}$, $10^{th}$, and $20^{th}$ mode, together with the corresponding Husimi distributions. Note that Husimi distributions coincide with the regions of phase space with the largest density of points. This agreement suggests that the TPM can be a used to predict, at last qualitatively, the Husimi distributions.\\

\section{Purely Quantum Features}

Recently, Ketzmerick and co-workers \cite{ketz2} obtained important results concerning the transport properties of waveguides which classically produce fully chaotic or mixed dynamics. Specifically, they showed that the behavior of the Landauer conductance $G$ (given by $\frac{2e}{\hbar^2}\sum_{n} \sum_{m}\mid t_{n,m} \mid^2$, where $t_{n,m}$ are the transmission elements of the $S$ matrix) and the Wigner delay time can clearly distinguish between full and mixed chaos. See also \cite{Akguc}. In our geometry, this distinction is exemplified in Fig. 12 where we contrast the dimensionless conductance $G$, in a range of energy that supports $20$ modes, for the cases of mixed (Fig. 12a) and globally chaotic dynamics (Fig. 12b). For the mixed case we used the same geometrical parameters as before [$(d,a,L)=(1,0.305,5.55)$] and for the globally chaotic case we use $(d,a,L)=(1,0.305,2.77)$. The difference is clear: while for mixed dynamics the conductance $G$ fluctuates strongly with sharp resonances, for global chaos it is a smooth function of energy. These wild fluctuations in the mixed chaos case are due to the existence of resonance and hierarchical states, as argued in \cite{ketz2}.\\

So far we have seen that the similarity between the classical SP matrix and its quantum counterpart enabled us to understand various quantum features in terms of purely classical dynamics (see also \cite{gala5}). We have also seen a nice agreement between the TPM and its quantum counterpart, the Husimi distributions. The energies for which such a good agreement occurs were chosen at random. However, if we select now an energy value corresponding to a sharp dip or peak in the conductance (mixed case) we find substancial differences between the classical and quantum quantities. As an example, in Fig. 13 we plot the same modes as in Fig. 11 but with an energy corresponding to a typical dip in the conductance, see inset of Fig. 12. Note that in contrast with Fig. 11 the Husimis lie predominantly on classically forbidden regions of phase space. In particular, for this energy they have their support on the four-period resonance. The culprit of this phenomenom is Heisenberg's uncertainty principle that allows the wavefunction to tunnel through the KAM barriers.\\
  
It is also instructive to look at the wavefunctions in configuration space for the non-resonant and resonant cases, see Fig. 14. The plots on the left (right) of this figure correspond to the non-resonant (resonant) case. The non-resonant (resonant) energy is the same as that used in Fig. 11 (Fig. 13) for modes 2, 5 and 20. These plots reveal a striking difference. In the resonant case we notice a ``bow-tie'' pattern which shadows the classical trajectory of a particle in a four-period periodic orbit, corresponding to the four islands in the TPM. As mentioned above, the phase space region of the period four islands is not accessible to {\it classical} particle incoming from the right or left leads; only trajectories originated in the interior of the cavity can be trapped and form this pattern. It is precisely this bow-tie pattern that has been exploited experimentally in \cite{Stone2} for the construction of high gain microlasers with directional emission with {\it closed} resonators of high refractive index. We see that such a pattern can be obtained also for {\it open} cavities because quantum mechanically the wavefunction, due to Heisenberg's uncertainty, can penetrate into the classically forbidben areas. Thus, as we have proposed in \cite{jamb1}, directional emission microlasers may also be constructed with open cavities. We have also computed ``classical wavefunctions'' for these type of cavities, constructed basically in the same way as the TPM with a box-counting method to mimic intensity in configuration space \cite{jamb2}. The classical wavefuntions that we obtained look very much like the truly quantum wave functions for the non-resonant case except for interference patterns. Of course, the classical wavefunctions do not reproduce the resonance structure as this is a purely quantum effect.\\

\section{Concluding Remarks}

The objective of this work has been to contribute to the understanding of the classical-quantum correspondence as regards scattering in open billiards which serve as models of mesoscopic electron waveguides as well as microwave cavities. Here we make a comparative analysis betweeen the Scattering Probability (SP) matrix and the Husimi distributions with their classical counterparts, namely  the classical SP matrix and the Transient Poincar\`{e} Maps (TPM), respectively. As a paradigm of open billiards, we use a model of a ballistic 2D waveguide formed by a rippled cavity attached to two colinear leads. The classical particle motion in such cavity is known to undergo the generic (Hamiltonian) transition to chaos \cite{Licht,gala1,ketz2,Month,gala2,gala3,gala4,Akguc}. For the purpose of this work, the parameters of the cavity were chosen to produce mixed chaos (a ternary incomplete horshoe \cite{Jung}).\\

We find a very good global similarity between classical and quantum SP matrices and, as expected, this similarity is greater as the number of open modes in the quantum system increases. However, even for a moderate number of modes ($\sim 10-20$) the similarity between classical and quantum SP matrices allows to extract important information about general as well as individual features of the quantum system. In particular, all high intensity patterns in the classical SP matrix, found to be produced by bundles of stable trajectories, appear also in the quantum SP matrix. Hence, knowledge of the structure of the {\it classical} SP matrix allows us to predict which incoming modes contribute dominantly to transmission or reflection in the quantum system \cite{gala5}. We have also shown that certain self-similar structure in the transmission part of the classical SP matrix is due to Whispering Gallery Orbits (WGO).\\

On the other hand, important differences are revealed for energy values corresponding to sharp peaks or dips in the conductance. For these resonance energies, the wavefunctions tunnel (thanks to Heisenberg's uncertainty principle) into forbidden classical phase space regions (resonance islands), associated with stable periodic orbits within the cavity.\\
 
Similarly, the Husimi distributions and the TPM show an excellent agreement in the case of non-resonant energies. The agreement suggests the usefulness of the TPM as a tool to predict the support of Husimi distributions for each open mode in the system. Also, we have seen for a certain resonant energy, where substantial differences between TPM and Husimi distributions occur, the wavefunctions of the system form a bow-tie pattern. We have exploited this phenomenom to propose the construction of microlaser resonators in open cavities \cite{jamb1}.\\

\vspace{1cm}

{\bf Acknowledgementes}: The authors thank B. Huckestein for the calculation of the $S$ matrix used to construct Fig. 3. JAMB wishes to thank the University Hradec Kralove and the Institute of Physics, Czech Academy of Sciences for they kind hospitality. We wish to acknowledge financial support from CONACyT (Mexico) grant No. 26163-E, and from project II-63G01, VIEP, BUAP.

\newpage

\begin{center}
{\Large \bf FIGURE CAPTIONS}
\end{center}
\vspace{.5cm}

{\bf Fig. 1} Geometry of the waveguide for a two-period rippled cavity. $\xi(x)=1-\cos(2\pi x/L)$.\\

{\bf Fig. 2} Poincar\`{e} Map for $(d,a,L)=(1,0.305,5.55)$. The Poincar\`{e} Surface of Section is located at $y=0$.\\

{\bf Fig. 3} Quantum SPM, $|S_{n,m}|^2$, for the one-period waveguide with 
$(d,a,L)=(1,0.305,5.55)$, and $M=200$.\\

{\bf Fig. 4} Classical SPM, $|S_{n,m}|^2$, for the one-period waveguide with 
$(d,a,L)=(1,0.305,5.55)$, and $M=200$.\\

{\bf Fig. 5} Transmission part of the classical SPM, $|t_{n,m}|^2$ as a 
function of angles.\\

{\bf Fig. 6} Enlargement of the transmission part of the CSM for 
$0.345 \leq \theta \leq 0.48$.\\

{\bf Fig. 7} Trajectories that contribute to the first six generations of the structure of Fig. 6 (from only one kick on the top boundary to six kicks 
before leaving the concave part of that boundary).\\

{\bf Fig. 8} Typical trajectories that contribute to the zones marked in 
Fig. 5.\\

{\bf Fig. 9} (a) Transient Poincar\`{e} Map (TPM) generated by particles entering to the cavity of the waveguide 
from the left lead. (b) TPM generated by particles from both leads. In both cases the Poincar\`{e} Surface of Section is located at $y=0$.\\

{\bf Fig. 10} Transient Poincar\`{e} Map with Poincar\`{e} Surface of Section  at the top boundary and the location (highlighted) of the WGO that contribute to the structure of Fig. 6.\\

{\bf Fig. 11} Fraction of the Transient Poincar\`{e} Map with Poincar\`{e} Surface of Section at $y=0$ for particles from the left lead (black dots) and the correspondig Husimi distribution (gray-white) for the (a) $2^{nd}$, (b) $5^{th}$, (c) $10^{th}$, and (d) $20^{th}$.\\

{\bf Fig. 12} Dimensionless conductance {\it G} for the range of energy that 
supports $20$ modes for the (a) mixed case, $(d,a,L)=(1,0.305,5.55)$; and (b) 
chaotic case, $(d,a,L)=(1,0.305,2.77)$. The one-period waveguide is used.\\

{\bf Fig. 13} Fraction of the Transient Poincar\`{e} Map with Poincar\`{e} Surface of Section at $y=0$ for particles from the left lead (black dots) and the correspondig Husimi distribution (gray-white) for the (a) $2^{nd}$, (b) $5^{th}$, (c) $10^{th}$, and (d) $20^{th}$; the resonant energy of the inset of Fig. 12 is used.\\

{\bf Fig. 14} Wavefuntions for the modes (a) $2^{nd}$, (b) $5^{th}$, and 
$20^{th}$ for a non-resonant (left column) and the resonant energy of the inset of Fig. 12 (right column).\\

\end{document}